\documentclass[twocolumn,showpacs,preprintnumbers]{revtex4}
\usepackage{amssymb}
\usepackage{amsmath}
\usepackage{graphicx}
\usepackage{dcolumn}
\usepackage{bm}

\input{tcilatex}

\begin{document}

\title{Theoretical stopping power of copper for protons using the shellwise
local plasma approximation}
\author{C. C. Montanari}
\email{mclaudia@iafe.uba.ar}
\author{J. E. Miraglia}
\affiliation{Instituto de Astronom\'{\i}a y F\'{\i}sica del Espacio, casilla de correo
67, sucursal 28, C1428EGA Buenos Aires, Argentina.\\
Departamento de F\'{\i}sica, Facultad de Ciencias Exactas y Naturales,
Universidad de Buenos Aires, Buenos Aires, Argentina.}
\date{\today }

\begin{abstract}
We present a theoretical study on the energy loss by protons in solid
copper. The \textit{ab initio} shellwise local plasma approximation (SLPA)
is employed for the inner-shells, and the Mermin dielectric function for the
valence electrons. The partial contribution of each sub-shell of target
electrons is calculated separately, including the screening among the
electrons of the same binding energy. Present results are obtained using the
SLPA and the known Hartree-Fock wave functions for copper. The theoretical
curves are compared with the experimental data available and with the
semi-empirical \textsc{srim08} values, showing very good agreement in a
large energy region (5 keV-30 MeV).
\end{abstract}

\maketitle

%PACS numbers: 34.50.Bw

\section{Introduction}

The stopping power of ions in solids is a necessary ingredient of many areas
of basic science and material technology \cite{paul07,ICRU73}. There are
numerous detailed models to describe the response of outer electrons as a
free electron gas (FEG) \cite{echenique90,arista02,salin}. In many cases the
stopping due to the FEG is enough for energies reaching up to the maximum of
the stopping. But for targets with the full $d$ or even $f$ subshells, the
bound electron contribution plays an important role even at intermediate
energies. The description of bound electrons from first principles is a
heavy task, especially for multielectronic atoms (see, for example the
binary collisional formalism in Refs. \cite{grande,sigmund00}).

In this work we describe the bound electron contribution to the stopping
power by employing a many electron model, the shellwise local plasma
approximation (SLPA). This is based on the local plasma approximation (LPA)
\cite{LPA1,Bonderup, Chu} to deal with bound electrons as an inhomogeneous
free electron gas. We apply this model within the dielectric formalism \cite%
{kwei88,wang94,fuhr} but assuming independent shell approximation \cite%
{archubi}. This means that in the SLPA the electrons of the same binding
energy respond to the ion passage as a whole (collectively), screening the
interaction with the impinging ion. This binding energy (or ionization gap)
is included explicitly by employing Levine and Louie dielectric function
\cite{levine}.

The SLPA allows us to calculate the different moments of the energy loss by
the ion when at least one of the bound electrons is ionized. This is an
\textit{ab initio} calculation -- no parameter is included -- whose only
inputs are the atomic densities of the different subshells and the
corresponding binding energies. Once the electronic wave functions are
available, the SLPA applies to the different targets with the same degree of
complexity. This model has been applied successfully to describe
experimental stopping cross sections of elements with $Z\leq 54$ (using the
Hartree-Fock wave functions and binding energies) \cite{LiZn}, and also for
very heavy elements like Au, Pb, and Bi for which numerical solutions of the
Dirac equation were needed \cite{relat}.

In what follows we present our theoretical results for the stopping of
protons in Cu (Z=29, [Ar] $3d^{10}4s^{1}$). The aim of this contribution is
to use Cu as a testing case for the SLPA because it is a well known element
(experimentally \cite{paul,srim}\ and theoretically \cite%
{sabin89,bichsel90,denton}), and because we can employ the known
Hartree-Fock wave functions \cite{bunge,clementi} to obtain the subshell
densities to include in the SLPA.

We organize this work in two steps. First, we describe briefly the
theoretical calculation, and then we present results for stopping cross
sections of Cu for protons and compare our \textit{ab initio} theoretical
values with the experimental data available \cite{paul} and with the
semi-empirical \textsc{srim08} results \cite{srim}.

\section{Theoretical calculations}

The SLPA\ introduces two important differences to the known LPA. One of them
is the shell--to--shell calculation. Mathematically it implies a separate
dielectric function for each subshell of bound electrons, and the total
response as the addition of these independent contributions. Physically, the
electrons are screened only by those of the same binding energy and not by
the rest.

The second one is the explicit inclusion of the binding energy by using the
Levine-Louie dielectric response \cite{levine}\ instead of Lindhard one \cite%
{lindhard}. This dielectric function takes into account the ionization gap
so that the contribution is null if the ionization threshold is deeper than
the energy transferred \cite{archubi}. The Levine-Louie dielectric response
maintains the characteristics of Lindhard \cite{lindhard} (linear response,
electron-electron correlation), and satisfies the $f$-sum rule (particle
number conservation). More detail about the SLPA can be found in Refs. \cite%
{LiZn,archubi}.

For a bare ion of charge $Z_{P}$\ and velocity $v$, the stopping cross
section due to the ionization of the $nl$ subshell of target electrons, is
expressed as

\begin{equation}
S_{q}^{nl}=\frac{2Z_{P}^{2}}{\pi v^{2}}\int_{0}^{\infty }\frac{dk}{k}%
\int_{0}^{kv}\omega \;\func{Im}[\frac{-1}{\varepsilon _{nl}(k,\omega )}]\
d\omega ,
\end{equation}

The dielectric function $\varepsilon _{nl}(k,\omega )$\ is calculated as a
mean value of a local response\textbf{\ }\cite{fuhr}
\begin{equation}
\func{Im}[\frac{-1}{\varepsilon _{nl}(k,\omega )}]=4\pi \int_{0}^{R_{WS}}%
\func{Im}[\frac{-1}{\varepsilon ^{LL}(k,\omega ,\delta _{nl}(r),E_{nl})}]\
r^{2}dr,  \label{epsLPA}
\end{equation}%
where $\delta _{nl}(r)$\ and $E_{nl}$ are the local density of electrons $%
\delta _{nl}(r)$ and ionization gap, respectively, obtained from the atomic
Hartree-Fock tabulated by Bunge \textit{et al.} \cite{bunge}. The spatial
integration is performed over the atomic dimensions within the solid Cu
(i.e., $R_{WS}$\ is the atomic Wigner-Seitz radius, related to the atomic
density $\delta _{at}=[\frac{4}{3}\pi R_{WS}^{3}]^{-1}$).

The dielectric response, given by Eq. (\ref{epsLPA}), considers only the
electrons in the $nl$ subshell. In this way the contribution of each shell
of target electrons is obtained separately. An orbital version of the LPA
was introduced many years ago by Meltzer \textit{et al. }\cite{meltzer}
within the logarithmic high energy limit of the stopping number. Instead, in
the SLPA we calculate a dielectric response of each shell, and it is valid
even in the intermediate energy range.

The independent shell description in the SLPA considers explicitly the
energy threshold $E_{nl}$\ of each shell (ionization of inner shells). The
Levine and Louie model \cite{levine} defines the dielectric function as

\begin{equation}
Im\left[ \varepsilon ^{LL}(q,\omega ,k_{nl}^{F})\right] =\left\{
\begin{array}{cc}
\begin{array}{c}
Im[\varepsilon ^{L}(q,\omega _{g},k_{nl}^{F}]%
\end{array}
&
\begin{array}{c}
\omega >|\epsilon _{nl}|%
\end{array}
\\
0 & \omega <|\epsilon _{nl}|%
\end{array}%
\right.  \label{LL}
\end{equation}%
with $\omega _{g}=\sqrt{\omega ^{2}+\epsilon _{nl}^{2}}$ and $\varepsilon
^{L}(q,\omega ,k_{nl}^{F})$ being the usual Lindhard dielectric function
\cite{lindhard}. This modified SLPA has already been applied successfully in
recent calculations of stopping power \cite{LiZn,garcia}, energy loss
straggling \cite{straggling} in solids, and also in multiple ionization
cross sections of rare gases \cite{archubi}.

\section{Results}

In order to compare with the experimental data available we calculate total
stopping cross sections. These values are obtained as the addition of the
bound electron and the FEG contributions. The former calculated with the
SLPA formalism as described before. The latter by employing the dielectric
formalism with the Mermin dielectric function \cite{mermin}.

The characteristic plasmon frequency and width of the Cu FEG are $\omega
_{p}=0.703$ a.u. and $\gamma =0.950$ a.u. respectively. This implies a mean
value of electrons in the FEG $Ne=3.14$, and a Seitz radius $r_{S}=1.82$
a.u..These values were obtained from the optical data of energy loss
function \cite{palik} by considering only the first important peak. The \
number of electrons in the FEG is similar to the experimental value
recommended by Isaacson \cite{isaacson}. To keep the total number of
electrons, we considered Cu as [Ar] $3d^{7.86}$ and $3.14$ electrons as FEG.

%\begin{figure}[tbp]
\begin{figure}[h!]
\begin{center}
\includegraphics*[width=8cm]{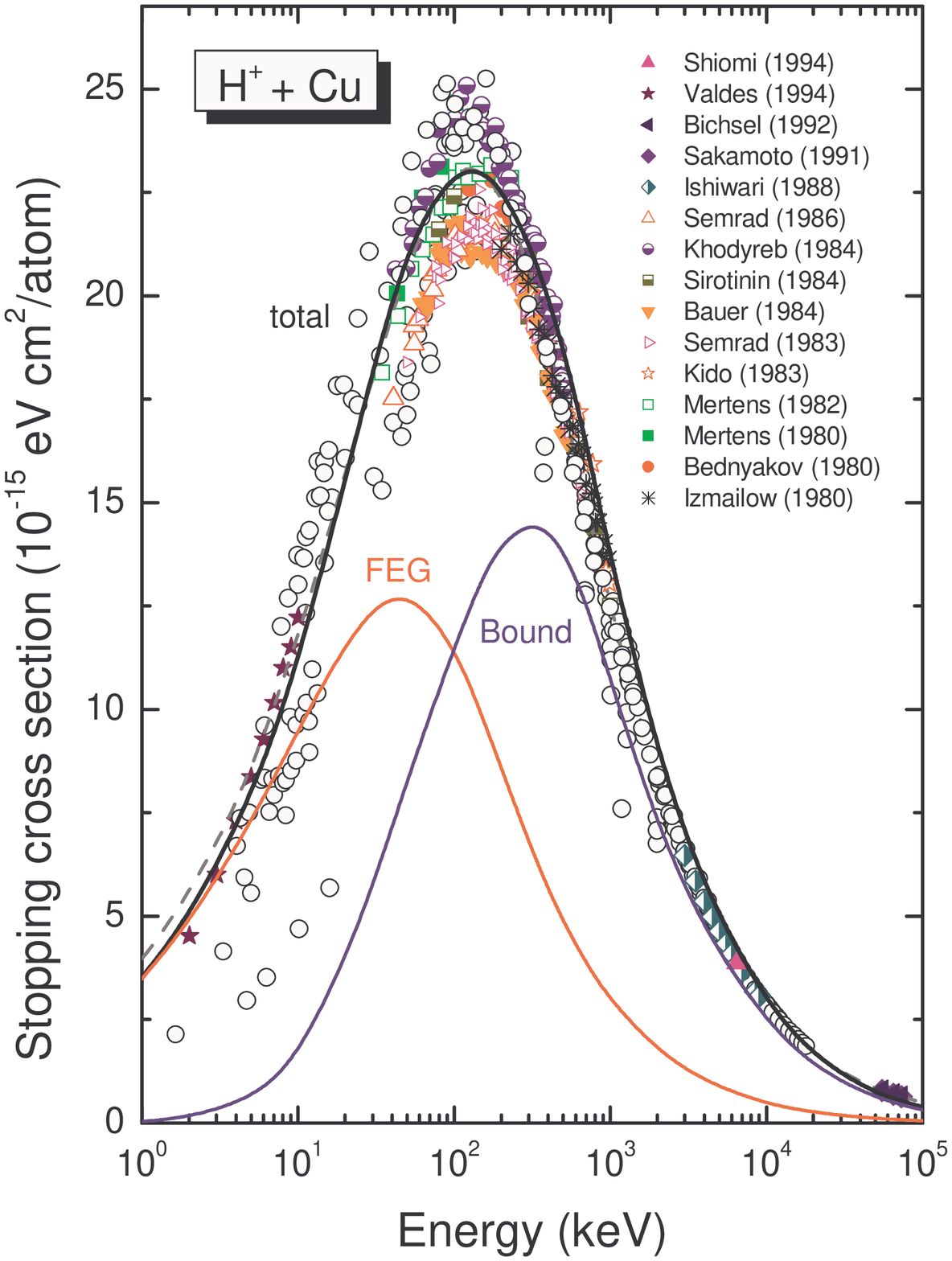}
\end{center}
\caption{Stopping cross sections of Cu for protons. Curves:
Solid--lines, present theoretical results for the contributions by
bound electrons (SLPA) and the FEG, and total stopping as the
addition of the previous two; dash--line, \textsc{srim08} results
\protect\cite{srim}. Symbols: black
hollow-circles, sets of experimental data compiled by H. Paul \protect\cite%
{paul} corresponding to (1935-1979); different symbols signed
within the figure, experimental data in Refs. \protect\cite{Si84,
Shiomi,vs94,Bh92,Sk91,is88,Se86,kh84,Br84a,se83,ki83,me82,me80,be80,iz80}.}
\label{HCu}%...............
\end{figure}
%................

In Fig. \ref{HCu} we plot our total stopping cross sections of Cu for
protons, and compare them with the experimental data available \cite{paul},
and with \textsc{srim08} results \cite{srim}. The contribution from the FEG
and the bound electrons are displayed separately. About the experimental
data, we indicate separately only the data since 1980, while previous one is
plotted together with a single type of symbol. The agreement with the
experimental data and with the \textsc{srim08} curve is good. This is a
perturbative formalism, which is not expected to describe the data for ion
energies below 30 keV. However, we have obtained a very good agreement even
with the measurements by Valdes \textit{et al.} \cite{vs94}\ in the low
energy region. The FEG results obtained with the Mermin dielectric function,
even perturbative, are rather good for protons in Cu at such low energies.
Figure \ref{HCu} shows that the stopping maximum is correctly described in
energy and value. For energies above that of the stopping maximum, the
description is good, but shows an overestimation around 2 MeV. This behavior
of the SLPA has already been noted in previous works \cite{relat,W}\ and it
is under study.

\section{Concluding remarks}

In this work we test the\textit{\ ab initio} SLPA for the stopping of Cu for
protons. This is a perturbative model which works within the dielectric
formalism by considering the response of bound electrons as that of a free
electron gas of inhomogeneous density. The SLPA introduces the independent
shell approximation to the usual formalism. The theoretical results are
compared with the experimental data available and with the semi empirical
\textsc{srim08} values, showing good agreement in a large energy region (5
keV-30 MeV).

\acknowledgments This work was partially supported by the Consejo Nacional
de Investigaciones Cient\'{\i}ficas y T\'{e}cnicas (CONICET), the
Universidad de Buenos Aires, and the Agencia Nacional de Promoci\'{o}n Cient%
\'{\i}fica y Tecnol\'{o}gica of Argentina.

%***************************************************************************

\end{document}